\documentstyle[prl,aps,preprint]{revtex}

\begin{document}

\bigskip\ 

\begin{center}
{\bf HURWITZ THEOREM AND PARALLELIZABLE SPHERES}

{\bf FROM TENSOR ANALYSIS}

\smallskip

\bigskip \ 

\bigskip \ 

J. A. Nieto\footnote{%
nieto@uas.uasnet.mx}and L. N. Alejo-Armenta\footnote{%
nabor@uas.uasnet.mx}

\smallskip

{\it Facultad de Ciencias F\'{\i}sico-Matem\'{a}ticas, Universidad
Aut\'{o}noma}

{\it de Sinaloa, C.P. 80010, Culiac\'{a}n Sinaloa, M\'{e}xico}

\smallskip

\bigskip

\bigskip

\bigskip \ 

{\bf Abstract}
\end{center}

By using tensor analysis, we find a connection between normed algebras and
the parallelizability of the spheres S$^1$, S$^3$ and S$^7.$ In this
process, we discovered the analogue of Hurwitz theorem for curved spaces and
a geometrical unified formalism for the metric and the torsion. In order to
achieve these goals we first develope a proof of Hurwitz theorem based in
tensor analysis. It turns out that in contrast to the doubling procedure and
Clifford algebra mechanism, our proof is entirely based in tensor algebra
applied to the normed algebra condition. From the tersor analysis point of
view our proof is straightforward and short. We also discuss a possible
connection between our formalism and the Cayley-Dickson algebras and Hopf
maps.

\smallskip\ 

\bigskip

\bigskip

\bigskip

\bigskip

Pacs No.: 02.10. Vr; 02.10. Tq; 02.90.+p.

January, 2001

\newpage

\noindent {\bf I. INTRODUCTION}

\smallskip\ 

It is known that normed algebras are closely related to supersymmetry$^{1-3}$
and super p-branes$^4$, and that these two theories require tensor analysis
for their formulation. Therefore, it may be interesting to study normed
algebras from the tensor analysis point of view. Moreover, normed algebras,
among other things, are physically interesting because they are division
algebras and in this context there are a number of interesting connections
with fundamental physics. Let us just give some few examples about this
fact. It has been shown$^5$ that a generalized instantons in eight
dimensions fit inside the family of gauge-theoretical solitons associated to
normed algebras. There is a deep relation between division algebras and
superparticles (see ref. 6, 7 and references there in) and twistor
formulation of a massless particles$^{8,9}$. Finite Lorentz transformations
of vectors in 10-dimensional Minkowski space have been studied$^{10}$ by
means of division algebras. Finally, division algebras seem to be deeply
related to the geometric structures of M-theory$^{11}$.

In this work, we show that tensor analysis can be used to give a
straightforward connection between normed algebras and the paralellizability
of the spheres S$^1$, S$^3$ and S$^7.$ In the process of studing this
connection, we discovered the analogue of Hurwitz theorem for curved spaces
and a unified formalism for the metric and the torsion. Our strategy to
achieve these goals was first to develope a proof of Hurwitz's theorem$^{12}$
based in tensor analysis. It turns out that this proof is essentially based
on the composition law rewritten in tensor notation. From the point of view
of tensor analysis, such a proof is short and straightforward. In fact, we
do not even require to use the doubling procedure$^{12}$ or the Clifford
algebra mechanism$^{13}$.

The plan of the article is as follows. In section II, we introduce tensor
notation and a proof of Hurwitz theorem based in tensor analysis. In section
III, we briefly review the Cartan-Shouten equations as presented by Gursey
and Tze. In section IV, using the Gursey-Tze's procedure, we show a
connection between our proof of Hurwitz theorem and the paralellizability of
the spheres S$^1$, S$^3$ and S$^7.$ We also prove that such a connection
leads to a generalization of Hurwitz theorem for curved spaces. In section
V, we develope a unified formalism for the metric and the torsion. Finally,
in section VI, we make a number of final comments and briefly outline a
possible extension of the present work to the case of Cayley-Dickson
algebras and Hopf maps.

\smallskip\ 

\noindent {\bf II. AN ALTERNATIVE PROOF OF HURWITZ THEOREM}

\smallskip\ 

Let us start recalling the Hurwitz theorem:

{\bf Theorem (Hurwitz, 1898): }{\it Every normed algebra with an identity is
isomorphic to one of following four algebras: the real numbers, the complex
numbers, the quaternions, and the Cayley (octonion) numbers.}

{\it Proof (alternative)}: Consider a $N=d+1$dimensional algebra ${\cal A}$
over the real numbers $R${\it .} Let

\begin{equation}
e_0,e_1,...,e_d  \label{ec.1}
\end{equation}
be a basis of ${\cal A}$, and let

\begin{equation}
A=A^0e_0+A^1e_1+...+A^de_d  \label{ec.2}
\end{equation}
be the representation of a vector $A$ $\epsilon $ ${\cal A}$ relative to
this basis. Here, $A^0,A^1,...,A^d\epsilon R.$ Take the multiplication table
in the form

\begin{equation}
\begin{array}{c}
e_ie_j=C_{ij}^0e_0+C_{ij}^1e_1+...+C_{ij}^de_d, \\ 
\\ 
(i,j=0,1,...,d),
\end{array}
\label{ec.3}
\end{equation}
where $C_{ij}^k$ , the so-called structure constants, are real numbers (See,
for instance, I. L. Kantor and A.S. Solodovnikov$^{12}$, S. Okubo$^{13}$,
Abdel-Khalek$^{14}$, J. Adem$^{15}$, F. R. Cohen$^{16}$, Y. A. Drozd and V.
V. Kirichenko$^{17}$.)

Assume that the basis (1) is orthonormal with bi-linear symmetric
non-degenerate scalar product given by

\begin{equation}
<e_i\mid e_j>=\delta _{ij},  \label{ec.4}
\end{equation}
where $\delta _{ij}$ is the so-called Kronecker delta, with $\delta _{ij}=0$
if $i\neq j$ and $\delta _{ij}=1$ if $i=j$ .

Assume the Einstein summation convention: if the same index appears twice,
once as superscript and once as a subscript, then the index is summed over
all possible values. This convention allows to write (2) and (3) as

\begin{equation}
A=A^ie_i,  \label{ec.5}
\end{equation}
and

\begin{equation}
e_ie_j=C_{ij}^ke_k,  \label{ec.6}
\end{equation}
respectively.

We shall assume that $e_i$ transforms as covariant first-rank tensor

\begin{equation}
e_i^{\prime }=\Lambda _i^je_j,  \label{ec.7}
\end{equation}
where, in order to leave invariant (4), $\Lambda _i^j$ satisfies the
conditions $\det \Lambda _i^j\neq 0$ and $\Lambda _k^i\Lambda _l^j\delta
_{ij}=\delta _{kl}$ and therefore $\Lambda _i^j$ is an element of an
orthogonal transformation $O(N)$. Since $A$ is an invariant quantity, from
(5) and (7) we find that $A^i$ should transform as

\begin{equation}
A^{\prime i}=\Lambda _j^iA^j,  \label{ec.8}
\end{equation}
i.e. $A^i$ is a contravariant first-rank tensor. While from (6) and (7) we
find that $C_{ij}^k$ transforms as 
\begin{equation}
C_{st}^{\prime r}=\Lambda _k^r\Lambda _s^i\Lambda _t^jC_{ij}^k,  \label{ec.9}
\end{equation}
i.e. $C_{ij}^k$ is a mixed third-rank tensor (twice covariant and once
contravariant).

According to the multiplication table (6) the product $AB=D$ for $A,B$ and $%
D\epsilon $ ${\cal A}$ is given by

\begin{equation}
A^iB^jC_{ij}^k=D^k.  \label{ec.10}
\end{equation}
A normed algebra is an algebra in which the composition law

\begin{equation}
<AB\mid AB>=<A\mid A><B\mid B>  \label{ec.11}
\end{equation}
holds for any $A$, $B$ $\epsilon $ ${\cal A}$. It can be shown that this
expression is equivalent to (see, for instance, section 3.1 of ref. 13)

\begin{equation}
<AB\mid CD>+<CB\mid AD>=2<A\mid C><B\mid D>,  \label{ec.12}
\end{equation}
where $A$, $B,C,D$ $\epsilon $ ${\cal A}$. Choosing

\begin{equation}
A\rightarrow e_i,B\rightarrow e_j,C\rightarrow e_m\text{ and }D\rightarrow
e_{n\text{ }}  \label{ec.13}
\end{equation}
we find that (12) leads to

\begin{equation}
<e_ie_j\mid e_me_n>+<e_me_j\mid e_ie_n>=2<e_i\mid e_m><e_j\mid e_n>.
\label{ec.14}
\end{equation}

Using (4) and (6), from (14) we obtain the key formula

\begin{equation}
C_{ij}^kC_{mn}^l\delta _{kl}+C_{mj}^kC_{in}^l\delta _{kl}=2\delta
_{im}\delta _{jn}.  \label{ec.15}
\end{equation}
Note that, although at first sight it looks like, (15) is not a Clifford
algebra. The reason for this is that, at this level, there are not any
symmetries between the indices $i,j$ and $k$ of $C_{ij}^k.$ In this work,
the formula (15) shall play a central role. Note that when $D=1,$ this
equation admits the solution $C_{00}^0=1.$ Therefore, in what follows we
shall be mainly interested in solutions of (15) when $D\neq 1.$

Let $e_0$ be the identity of the algebra ${\cal A}{\it .}$ Then, the
multiplication table (6) implies

\begin{equation}
e_0e_j=C_{0j}^ke_k=e_j  \label{ec.16}
\end{equation}
and

\begin{equation}
e_je_0=C_{j0}^ke_k=e_j.  \label{ec.17}
\end{equation}
From (16) we find

\begin{equation}
C_{0j}^k=\delta _j^k,  \label{ec.18}
\end{equation}
while from (17) we obtain

\begin{equation}
C_{j0}^k=\delta _j^k,  \label{ec.19}
\end{equation}
where $\delta _j^k$ is also a Kronecker delta.

Let us now split the formula (15) as follows:

\begin{equation}
C_{0j}^kC_{0n}^l\delta _{kl}+C_{0j}^kC_{0n}^l\delta _{kl}=2\delta _{jn},
\label{ec.20}
\end{equation}

\begin{equation}
C_{i0}^kC_{m0}^l\delta _{kl}+C_{m0}^kC_{i0}^l\delta _{kl}=2\delta _{im},
\label{ec.21}
\end{equation}

\begin{equation}
C_{0j}^kC_{an}^l\delta _{kl}+C_{aj}^kC_{0n}^l\delta _{kl}=0,  \label{ec.22}
\end{equation}

\begin{equation}
C_{i0}^kC_{ma}^l\delta _{kl}+C_{m0}^kC_{ia}^l\delta _{kl}=0,  \label{ec.23}
\end{equation}

\begin{equation}
C_{ab}^0C_{cd}^0+C_{cb}^0C_{ad}^0+C_{ab}^eC_{cd}^f\delta
_{ef}+C_{cb}^eC_{ad}^f\delta _{ef}=2\delta _{ac}\delta _{bd},  \label{ec.24}
\end{equation}
where the indices $a,b,$..., etc run from $1$ to $d$. Using (18) and (19) we
note that the equations (20) and (21) are identities. Moreover, the
expression (22) gives

\begin{equation}
C_{anj}+C_{ajn}=0,  \label{ec.25}
\end{equation}
while (23) leads to

\begin{equation}
C_{mai}+C_{iam}=0,  \label{ec.26}
\end{equation}
where $C_{mai}=C_{ma}^l\delta _{il},$ i.e. we raised and lowed indices with $%
\delta ^{il}$ and $\delta _{il}$ respectively. From (25) we obtain

\begin{equation}
C_{ab0}+C_{a0b}=0,  \label{ec.27}
\end{equation}
and

\begin{equation}
C_{abc}+C_{acb}=0.  \label{ec.28}
\end{equation}
While from (26) we get 
\begin{equation}
C_{ba0}+C_{0ab}=0  \label{ec.29}
\end{equation}
and

\begin{equation}
C_{bac}+C_{cab}=0.  \label{ec.30}
\end{equation}
\ Thus, using (18) and (19), we have that either (27) or (30) implies that

\begin{equation}
C_{ab}^0=-\delta _{ab},  \label{ec.31}
\end{equation}
while (28) and (30) mean that the quantity $C_{abc}$ is completely
antisymmetric.

Now, by substituting (31) into (24) we obtain

\begin{equation}
C_{ab}^eC_{cd}^f\delta _{ef}+C_{cb}^eC_{ad}^f\delta _{ef}=2\delta
_{ac}\delta _{bd}-\delta _{ab}\delta _{cd}-\delta _{cb}\delta _{ad}.
\label{ec.32}
\end{equation}
Multiplying this equation by $\delta ^{ac}$ and $C_g^{ad}$ $=\delta
^{ae}\delta ^{af}C_{gef}$ we find

\begin{equation}
\delta ^{ac}C_{ab}^eC_{cd}^f\delta _{ef}=(d-1)\delta _{bd}  \label{ec.33}
\end{equation}
and

\begin{equation}
C_g^{ad}C_{ab}^eC_{cd}^f\delta _{ef}+C_g^{ad}C_{cb}^eC_{ad}^f\delta
_{ef}=3C_{gcb},  \label{ec.34}
\end{equation}
respectively, where we used the fact that $C_{abc}$ is completely
antisymmetric. Moreover, using again the property that $C_{abc}$ is
completely antisymmetric, we find that (33) becomes

\begin{equation}
C_b^{ce}C_{dce}^{\ }=(d-1)\delta _{bd},  \label{ec.35}
\end{equation}
while substituting (33) into (34) we have

\begin{equation}
C_{dg}^aC_{ab}^eC_{ec}^d=(d-4)C_{gbc}.  \label{ec.36}
\end{equation}
Multiplying (35) by $\delta ^{bd}$ we find the formula

\begin{equation}
C^{abc}C_{abc}=d(d-1),  \label{ec.37}
\end{equation}
which for $d=0$ and $d=1,$ admits the solution $C_{abc}=0$. Moreover, for $%
d=3$ the formula (37) admits the solution $C_{abc}^{\text{ }}=\varepsilon
_{abc},$ where $\varepsilon _{abc}$ is the completely antisymmetric
Levi-Civita symbol, with $\varepsilon _{123}=1.$

Let us define

\begin{equation}
G_{abc}\equiv C_{da}^gC_{gb}^eC_{ec}^d.  \label{ec.38}
\end{equation}
Since $C_{abc}^{\ \text{ }}$ is completely antisymmetric, we find that $%
G_{abc}$ is also completely antisymmetric. From (36) and (38) we find that

\begin{equation}
G^{abc}G_{abc}=(d-4)^2C^{abc}C_{abc},  \label{ec.39}
\end{equation}
which by virtue of (37) leads to

\begin{equation}
G^{abc}G_{abc}=d(d-1)(d-4)^2.  \label{ec.40}
\end{equation}

Substituting (38) into (40) we get

\begin{equation}
C_h^{ag}C_g^{br}C_r^{ch}C_{da}^eC_{eb}^fC_{fc}^d=d(d-1)(d-4)^2,
\label{ec.41}
\end{equation}
which can be rewritten in the form

\begin{equation}
C_h^{ag}C_g^{br}C_{da}^eC_{eb}^f(C_r^{ch}C_{fc}^d)=d(d-1)(d-4)^2.
\label{ec.42}
\end{equation}
So, considering (32) we find that (42) becomes

\begin{equation}
C_h^{ag}C_g^{br}C_{da}^eC_{eb}^f(2\delta _{rf}\delta ^{hd}-\delta _r^h\delta
_f^d-\delta _r^d\delta _f^h-C_f^{ch}C_{rc}^d)=d(d-1)(d-4)^2.  \label{ec.43}
\end{equation}
Now, using (35), (36) and (37) and the fact that $C_{abc}$ is completely
antisymmetric, we obtain

\begin{equation}
\begin{array}{cc}
C_h^{ag}C_g^{br}C_{da}^eC_{eb}^f(2\delta _{rf}\delta ^{hd}-\delta _r^h\delta
_f^d-\delta _r^d\delta _f^h)= &  \\ 
&  \\ 
=2d(d-1)(d-1)-d(d-1)(d-1)+d(d-1)(d-4) &  \\ 
&  \\ 
=d(d-1)(2d-5), & 
\end{array}
\label{ec.44}
\end{equation}
while, since with respect to the indices $a$ and $h$ the quantity $C_h^{ag}$
is antisymmetric and the tensor $(C_{da}^eC_{eb}^fC_f^{ch}C_{rc}^d)$ is
symmetric, we get

\begin{equation}
C_h^{ag}C_g^{br}C_{da}^eC_{eb}^f(C_f^{ch}C_{rc}^d)=C_h^{ag}C_g^{br}(C_{da}^eC_{eb}^fC_f^{ch}C_{rc}^d)\equiv 0.
\label{ec.45}
\end{equation}
Thus, by substituting the results (44) and (45) into (43), we discover the
equation

\begin{equation}
d(d-1)(2d-5)=d(d-1)(d-4)^2,  \label{ec.46}
\end{equation}
which can be rewritten in the form

\begin{equation}
d(d-1)(d-3)(d-7)=0.  \label{ec.47}
\end{equation}
The only solutions for this equation are $d=0,1,3$ and $7.$ Therefore, we
have shown that the equation (15) has solution only for $D=1,2,4$ and $8.$
This implies that normed algebras with unit element are only possible in
these dimensions.

We have yet to show that the cases $D=1,D=2,D=4$ and $D=8$ correspond to
real, complex, quaternion and octonion algebras, respectively. The case $D=1$
is trivial since for any $A\epsilon {\cal A},$ we have $A=A^0e_0,$ where $%
A^0\epsilon $ $R$. For the case $D=2,$ we have $C_{abc}=0$, $%
C_{ab}^0=-\delta _{ab},$ $C_{n0}^s=\delta _n^s$ and $C_{0n}^s=\delta _n^s.$
These values of the structure constants determine the algebra of complex
numbers. While, for the case $D=4,$ we have the solution of (32) $%
C_{abc}=\varepsilon _{abc}$, $C_{ab}^0=-\delta _{ab},$ $C_{n0}^s=\delta _n^s$
and $C_{0n}^s=\delta _n^s$. It is well known that these values of the
structure constants determine the algebra of quaternions. Finally, for the
case $D=8$ we have $C_{ab}^0=-\delta _{ab},$ $C_{n0}^s=\delta _n^s$ and $%
C_{0n}^s=\delta _n^s.$ Now, take the structure constants as $C_{abc}=\Xi
_{abc}$, where $\Xi _{abc}$ is a completely antisymmetric Levi-Civita
symbol, with $\Xi _{abc}=1,$ for the following values of the indices $%
(a,b,c) $:

\begin{equation}
(1,2,3),(5,1,6),(6,2,4),(4,3,5),(1,7,4),(3,7,6)\text{ and }(2,7,5).
\label{ec.48}
\end{equation}
In fact, these values of the structure constants determine the algebra of
octonions. One can verify by straightforward, but tedious, computation that,
in fact for $d=7,$ these values for the structure constants give a solution
of (32)$.$

It is known that by definition two $(d+1)-$ dimensional algebras ${\cal A}%
^{\prime }$ and ${\cal A}$ are said to be isomorphic if they have bases with
identical multiplication table. Therefore, it remains to show that any other
solution is isomorphic to one of the above four solutions corresponding to
the real numbers, the complex numbers, the quaternions and the octonions.
For this purpose it is convenient to set $e_0^{\prime }=e_0.$ So that from
the transformation rule (7) we find that $\Lambda _0^0=1$ and $\Lambda
_0^a=0.$ Thus, from the relation $\Lambda _k^i\Lambda _l^j\delta
_{ij}=\delta _{kl},$ which leave invariant the scalar product (4), we find
that $\Lambda _a^0=0$ and therefore we have now the relation $\Lambda
_a^c\Lambda _b^d\delta _{cd}=\delta _{ab}$ which leaves invariant the scalar
product $<e_a\mid e_b>=\delta _{ab}.$ Consequently, we have that $\Lambda
_a^d$ is an element of $O(d)=$ $O(D-1)$ which is a subgroup of $O(D).$ Note
that the property $\det \Lambda _j^i\neq 0$ now becomes $\det \Lambda
_a^d\neq 0.$ Clearly, the transformation $\Lambda _a^d$ acts over elements
of the sub-vector space ${\cal A}_0$ of ${\cal A}$ defined by ${\cal A}%
_0=\{A\mid <A\mid e_0>=0,$ $A\epsilon {\cal A}\}$, with Dim ${\cal A}_0=d.$
In fact, we can write ${\cal A}$ $=\lambda e_0\oplus {\cal A}_0$, with $%
\lambda \epsilon R.$

Thus, we find that the structure constants $C_{abc}$ transform according to

\begin{equation}
C_{abc}^{^{\prime }}=\Lambda _a^d\Lambda _b^e\Lambda _c^fC_{def}.
\label{ec.49}
\end{equation}
Note that, since $\Lambda _a^d\Lambda _b^e\delta _{de}=\delta _{ab},$ if $%
C_{abc}$ is a solution of (32), then $C_{abc}^{^{\prime }}$ is also a
solution.

The transformation (49) has the important property that $C_{def}=0$ if and
only if $C_{abc}^{^{\prime }}=0.$ Therefore for real numbers, as well as for
complex numbers, the two algebras ${\cal A}^{\prime }$ and ${\cal A}$ are
isomorphic. For quaternions take $C_{def}=\varepsilon _{def}$ then (49)
implies that $C_{abc}^{^{\prime }}=\Lambda \varepsilon _{abc}$, $\Lambda
\equiv \det \Lambda _a^d.$ Thus, if $C_{def}=\varepsilon _{def}$ is a
solution of (32) we have that $C_{abc}^{^{\prime }}=\Lambda \varepsilon
_{abc}$ is also a solution. Therefore, for $D=4$ any solution of (32) is
isomorphic to the quaternionic solution, corresponding to $%
C_{abc}=\varepsilon _{abc}$. Similarly, for octonions applying (49) to the
completely antisymmetric symbol $\Xi _{abc}$ we find that $\Xi
_{abc}^{\prime }=\Lambda \Xi _{abc},$ where the values of the indices $%
(a,b,c)$ are given in (48).

Therefore, we have shown that up to isomorphism the dimensions $D=1,2,4$ and 
$8$ correspond to real, complex, quaternion and octonion algebras,
respectively. And in this way using the mathematical tool of tensor analysis
we have given an alternative proof of Hurwitz theorem. It is an interesting
and remarkable fact that without using doubling procedure (see ref. 12) or
Clifford algebra mechanism (see ref. 13) our proof has been based almost
completely in tensor algebra applied to the formula (15).

\smallskip\ 

\noindent {\bf III. CARTAN-SHOUTEN EQUATIONS}

\smallskip\ 

Define the metric tensor by

\begin{equation}
g_{ab}=\delta _{cd}h_a^{(c)}h_b^{(d)},  \label{ec.50}
\end{equation}
where $h_a^{(c)}$ = $h_a^{(c)}(x^{b)})$ is a vielbein field. Here, $x^a$ is
a coordinate patch of the geometrical sphere S$^d$. The quantities $C_{abc}$
can now be related to the S$^d$ torsion in the form

\begin{equation}
T_{abc}=r^{-1}C_{efg}h_a^{(e)}h_b^{(f)}h_c^{(g)},  \label{ec.51}
\end{equation}
where $r$ is the radius of S$^d.$ Using (35), (36), (50) and (51) we find
that the torsion $T_{abc}$ satisfies the equations:

\begin{equation}
T_a^{cd}T_{bcd}=(d-1)r^{-2}g_{ab},  \label{ec.52}
\end{equation}
and

\begin{equation}
T_{ea}^dT_{db}^fT_{fc}^e=(d-4)r^{-2}T_{abc.}  \label{ec.53}
\end{equation}
We recognize these expressions as the Cartan-Schouten equations$^{18}$which
as Gursey and Tze$^{19}$ noted, are mere septad-dressed, i.e. covariant
forms of the algebraic identities (35) and (36). It is well known that these
equations are closely related to the parallelizability of S$^1$, S$^3$ and S$%
^7$ (see ref. 13). In fact, the equations (52) and (53) can be derived by
adding to the riemannian symmetric connection $\Gamma _{ab}^c$ the totally
antisymmetric torsion tensor $T_{ab}^c$ and ''flattening'' the space in the
sense that

\begin{equation}
{\cal R}_{bcd}^{\ a}(\{\Omega _{ab}^c\})=0,  \label{ec.54}
\end{equation}
where

\begin{equation}
{\cal R}_{bcd}^{\ a}=\partial _c\Omega _{bd}^a-\partial _d\Omega
_{bc}^a+\Omega _{ec}^a\Omega _{bd}^e-\Omega _{ed}^a\Omega _{bc}^e,
\label{ec.55}
\end{equation}
with

\begin{equation}
\Omega _{ab}^c=\Gamma _{ab}^c+T_{ab}^c.  \label{ec.56}
\end{equation}
For our purpose it is convenient to show explicitly that in fact the
equations (52) and (53) follow from (54)-(56). By substituting (56) into
(54) we find

\begin{equation}
0=R_{bcd}^{\ a}+D_cT_{bd}^a-D_dT_{bc}^a+T_{ec}^aT_{bd}^e-T_{ed}^aT_{bc}^e,
\label{ec.57}
\end{equation}
Here, $D_c$ denotes a covariant derivative with $\Gamma _{ab}^c$ as a
connection and

\begin{equation}
R_{bcd}^{\ a}=\partial _c\Gamma _{bd}^a-\partial _d\Gamma _{bc}^a+\Gamma
_{ec}^a\Gamma _{bd}^e-\Gamma _{ed}^a\Gamma _{bc}^e.  \label{ec.58}
\end{equation}
Using in (57) the cyclic identities for $R_{\quad bcd}^{\ a}$ leads to

\begin{equation}
D_cT_{bda}=T_{e[bd}T_{a]c}^e,  \label{ec.59}
\end{equation}
where

\begin{equation}
T_{e[bd}T_{a]c}^e\equiv \frac 13%
\{T_{ebd}T_{ac}^e+T_{eab}T_{dc}^e+T_{eda}T_{bc}^e\}.  \label{ec.60}
\end{equation}
Substituting (59) into (57) we obtain

\begin{equation}
R_{abcd}^{\ }=T_{eab}T_{cd}^e-T_{e[ab}T_{c]d}^e.  \label{ec.61}
\end{equation}
For the sphere S$^d$ we have

\begin{equation}
R_{abcd}^{\ }=\frac 1{r^2}(g_{ac}g_{bd}-g_{ad}g_{bc}).  \label{ec.62}
\end{equation}
and therefore we get the equation

\begin{equation}
\frac 1{r^2}(g_{ac}g_{bd}-g_{ad}g_{bc})=T_{eab}T_{cd}^e-T_{e[ab}T_{c]d}^e.
\label{ec.63}
\end{equation}
Contracting in (63) with $g^{ac}$ leads to first Cartan-Shouten equation
(52), while contracting (63) with $T_f^{ac}$ leads to the second
Cartan-Shouten equation (53).

\smallskip\ 

\noindent {\bf IV. NORMED ALGEBRAS AND PARALLELIZABILITY OF }S$^1$, S$^3$
and S$^7$

\smallskip\ 

Let us start `undressing' (63). Using (50) and (51) we find

\begin{equation}
(\delta _{ac}\delta _{bd}-\delta _{ad}\delta
_{bc})=C_{eab}C_{cd}^e-C_{e[ab}C_{c]d}^e.  \label{ec.64}
\end{equation}
We shall show that this formula is equivalent to the formula (32). For this
purpose, let us rewrite formula (32) in form.

\begin{equation}
2\delta _{ac}\delta _{bd}-\delta _{ab}\delta _{cd}-\delta _{ad}\delta
_{cb}=C_{ab}^eC_{cd}^f\delta _{ef}+C_{cb}^eC_{ad}^f\delta _{ef}.
\label{ec.65}
\end{equation}
Let us first show that (64) implies (65). Making the change of indices $%
a\rightarrow c$ and $c\rightarrow a$ in (64) we find

\begin{equation}
(\delta _{ca}\delta _{bd}-\delta _{cd}\delta
_{ba})=C_{ecb}C_{ad}^e-C_{e[cb}C_{a]d}^e.  \label{ec.66}
\end{equation}
By adding (64) and (66) one easily obtains (65).

Let us now show that (65) implies (64). Let us start writing (65) in the form

\begin{equation}
C_{ab}^eC_{cd}^f\delta _{ef}-(\delta _{ac}\delta _{bd}-\delta _{ad}\delta
_{cb})+C_{cb}^eC_{ad}^f\delta _{ef}-(\delta _{ac}\delta _{bd}-\delta
_{ab}\delta _{cd})=0.  \label{ec.67}
\end{equation}
This expression suggests to define

\begin{equation}
F_{abcd}\equiv C_{ab}^eC_{cd}^f\delta _{ef}-(\delta _{ac}\delta _{bd}-\delta
_{ad}\delta _{cb}).  \label{ec.68}
\end{equation}
Therefore (67) gives

\begin{equation}
F_{abcd}+F_{cbad}=0.  \label{ec.69}
\end{equation}
Thus, considering that $C_{ab}^e$ is completely antisymmetric, from (68) and
(69) we discover that $F_{abcd}$ is also completely antisymmetric. Using
this important cyclic property for $F_{abcd}$ it is not difficult to show
that

\begin{equation}
F_{abcd}=C_{e[ab}C_{c]d}^e.  \label{ec.70}
\end{equation}
Substituting this result into (68) lead us back to (64). Thus, we have
proved the equivalence between (64) and (65).

With this equivalence at hand we have a number of interesting observations.
First, since in section II we showed that (65) (or (32)) admits solution
only for dimensions $d=1,3$ and $7$ we have that (64) admits solution only
in these dimensions. But, since (64) is the necessary and sufficient
condition for the existence of parallelism in S$^d,$ this means that we have
found an alternative proof of the fact that only the spheres S$^1,$S$^3$ and
S$^7$ are parallelizables. Second, in section II we proved that (65) is a
consequence of the normed condition (15) (or equivalent of (11)), while in
section III we proved that (64) is a consequence of the paralizability
condition (54). Therefore, we have find a new bridge between normed algebras
and parallelizable spheres. This link can be more transparent if using (50)
and (51) we dress (65) in the form

\begin{equation}
\frac 1{r^2}%
(2g_{ac}g_{bd}-g_{ab}g_{cd}-g_{ad}g_{cb})=T_{ab}^eT_{cd}^fg_{ef}+T_{cb}^eT_{ad}^fg_{ef}.
\label{ec.71}
\end{equation}
Of course, the equations (63) and (71) are equivalent. So, from (65) we can
derive (71) which in turn leads to the formula (63). Going backwards from
(63) we get (61). Therefore, we have shown that normed algebra condition
(65) implies the parallelizable condition (61). Similarly, we can show that
the parallelizable condition (61) implies the composition law (65).
Moreover, (31) and (71) suggest to define

\begin{equation}
T_{ab}^0\equiv -r^{-1}g_{ab}.  \label{ec.72}
\end{equation}
Thus, using (72) we find that (71) can be written in the form

\begin{equation}
T_{ab}^kT_{cd}^lg_{kl}+T_{cb}^kT_{ad}^lg_{kl}=\frac 2{r^2}g_{ac}g_{bd}.
\label{ec.73}
\end{equation}
where we recall that the indices $m$ and $n$ run from $0$ to $d.$ Setting $%
g_{00}=1$ and $g_{0a}=0$ we obtain (72) from (73)$.$ If we now take $%
T_{0j}^k=\delta _j^k$ and $T_{j0}^k=\delta _j^k,$ then we can generalize
(80) in the form

\begin{equation}
T_{ij}^kT_{mn}^lg_{kl}+T_{mj}^kT_{in}^lg_{kl}=\frac 2{r^2}g_{im}g_{jn}.
\label{ec.74}
\end{equation}
If we now introduce a basis $h_m$ such that

\begin{equation}
<h_m\mid h_n>=g_{mn},  \label{ec. 75}
\end{equation}
and

\begin{equation}
h_mh_n=T_{mn}^kh_k,  \label{ec. 76}
\end{equation}
we find that (74) leads to a generalization of (14)

\begin{equation}
<h_ih_j\mid h_mh_n>+<h_mh_j\mid h_ih_n>=\frac 2{r^2}<h_i\mid h_m><h_j\mid
h_n>.  \label{ec.77}
\end{equation}
Clearly, this expression implies the generalized composition law condition

\begin{equation}
<AB\mid AB>=\frac 1{r^2}<A\mid A><B\mid B>,  \label{ec.78}
\end{equation}
where $A=A^ih_i$.

The $r^{2\text{ }}$in the right hand side of (74) remind us that our
construction is valid for spheres. However, the equation (74) allows an
straightforward generalization. In fact, let us prove the theorem ($\star $)
below:

Before going into the details of the theorem let us define a `curved' space
as a space in which (75) and (74) hold, with $g_{00}=1,$ $g_{0a}=0$ and $%
g_{ab}=g_{ab}(x^i)=\eta _{cd}h_a^{(c)}(x^i)h_b^{(d)}(x^i),$ where the flat
metric $\eta _{cd}$ is diagonal and has an arbitrary signature and $%
T_{ij}^k=T_{ij}^k(x^i).$

{\bf Theorem (}$\star $){\bf : }{\it The possible dimensions }$D$ {\it of
any real normed algebra over a `curved' space with an identity are limited
to only 1, 2, 4 and 8.}

{\it Proof: }Let us write the composition law as follows: 
\begin{equation}
<h_ih_j\mid h_mh_n>+<h_mh_j\mid h_ih_n>=2<h_i\mid h_m><h_j\mid h_n>.
\label{ec.79}
\end{equation}
{\it \ }By virtue of (75) and (76) we find that (79) can be written as

\begin{equation}
T_{ij}^kT_{mn}^lg_{kl}+T_{mj}^kT_{in}^lg_{kl}=2g_{im}g_{jn}.  \label{ec.80}
\end{equation}
Taking $h_0$ as the identity with the properties that

\begin{equation}
<h_0\mid h_0>=g_{00}=1,  \label{ec. 81}
\end{equation}
and

\begin{equation}
<h_0\mid h_a>=g_{0a}=0,  \label{ec. 82}
\end{equation}
and following the same procedure as in section II, we find

\begin{equation}
T_{0j}^k=T_{j0}^k=\delta _j^k,  \label{ec. 83}
\end{equation}

\begin{equation}
T_{ab}^0=-g_{ab},  \label{ec.84}
\end{equation}

\begin{equation}
(2g_{ac}g_{bd}-g_{ab}g_{cd}-g_{ad}g_{cb})=T_{ab}^eT_{cd}^fg_{ef}+T_{cb}^eT_{ad}^fg_{ef}
\label{ec.85}
\end{equation}
with the property that $T_{ab}^e$ is completely antisymmetric$.$ The rest of
the story is similar to section II after formula (32). We find that (85) has
solution only if $d=1,3$ and $7.$ Note that in this result $g_{ab}$ may be
the metric not only for the spheres S$^1,$S$^3$ and S$^7,$ but also the
metric of any curved space. Moreover, in `flat' space $g_{ab}$ may have an
arbitrary signature. In particular for $D=4$ we could associate to $g_{ij}$
the signature ($g_{ij})=diag(1,1,1,-1)$ which correspond to Minkowski
signature. Note also that $T_{ij}^k$ unifies the metric $g_{ab}$ and the
torsion $T_{ab}^e$.

Summarizing, we have proved not only an equivalence between the Hurwitz
theorem for normed algebras and Cartan-Shouten theorem for parallelizable
spheres, but also the theorem ($\star $)$.$

\smallskip\ 

\noindent {\bf V. UNIFIED FORMALISM OF THE METRIC AND THE TORSION}

\smallskip\ 

In the previous section, in the context of normed algebras, we showed that
makes sense to unify the metric and the torsion in just one mathematical
object: the third-rank tensor $T_{ij}^k$ . A natural question is to see what
is the geometry induced by $T_{ij}^k.$ In this section we show that from the
vanishing of the Riemann tensor associated to such a third-rank tensor it
follows the metricity condition and the Cartan-Shouten equations for
homogeneous spacetimes.

Consider the equation

\begin{equation}
{\cal R}_{jkl}^{\ i}(\Omega _{jk}^i)=0,  \label{ec.86}
\end{equation}
where

\begin{equation}
{\cal R}_{jkl}^i=\partial _k\Omega _{jl}^i-\partial _l\Omega _{jk}^i+\Omega
_{mk}^i\Omega _{jl}^m-\Omega _{el}^i\Omega _{jk}^e  \label{ec.87}
\end{equation}
and

\begin{equation}
\Omega _{jk}^i=\Gamma _{jk}^i+T_{jk}^i.  \label{ec.88}
\end{equation}
These equations are, of course the analogue of the parallelizability
conditions (54)-(56). Let us see what are the consequences of (86)-(88). For
this purpose let us assume that $T_{jk}^i$ satisfies (83) and (84) and let
us set

\begin{equation}
\Gamma _{jk}^0=0\text{ and }\Gamma _{0k}^i=0.  \label{ec. 89}
\end{equation}
Thus, the non-vanishing terms of $\Omega _{jk}^i$ are

\begin{equation}
\Omega _{ab}^c=\Gamma _{ab}^c+T_{ab}^c,  \label{ec. 90}
\end{equation}

\begin{equation}
\Omega _{ab}^0=-g_{ab},  \label{ec. 91}
\end{equation}

\begin{equation}
\Omega _{0b}^a=\delta _b^a  \label{ec. 92}
\end{equation}
and

\begin{equation}
\Omega _{00}^0=1.  \label{ec. 93}
\end{equation}
At this stage it is important to note that (90)-(93) could also be obtained
if instead of (83), (84) and (89) we set $T_{jk}^0=0,T_{0k}^i=0,\Gamma
_{0k}^0=0$, $\Gamma _{00}^0=1,$ $\Gamma _{0b}^a=\delta _b^a$and $\Gamma
_{ab}^0=-g_{ab}.$ However, with this choice, the connection between (80) and
(85) will be lost. This connection is, of course, important to make contact
with the normed algebras for `curved' space discussed in the previous
section. It is worth mentioning that the formulae (83), (84) and (89) can be
understood as an anzats in the sense of Kaluza-Klein theory.

Let us split (87) in the form

\begin{equation}
{\cal R}_{abc}^0=\partial _b\Omega _{ac}^0-\partial _c\Omega _{ab}^0+\Omega
_{0b}^0\Omega _{ac}^0+\Omega _{db}^0\Omega _{ac}^d-\Omega _{0c}^0\Omega
_{ab}^0-\Omega _{dc}^0\Omega _{ab}^d,  \label{ec.94}
\end{equation}

\begin{equation}
{\cal R}_{a0c}^0=\partial _0\Omega _{ac}^0-\partial _c\Omega _{a0}^0+\Omega
_{00}^0\Omega _{ac}^0+\Omega _{d0}^0\Omega _{ac}^d-\Omega _{0c}^0\Omega
_{a0}^0-\Omega _{dc}^0\Omega _{a0}^d,  \label{ec.95}
\end{equation}

\begin{equation}
{\cal R}_{a0c}^b=\partial _0\Omega _{ac}^b-\partial _c\Omega _{a0}^b+\Omega
_{00}^b\Omega _{ac}^0+\Omega _{d0}^b\Omega _{ac}^d-\Omega _{0c}^b\Omega
_{a0}^0-\Omega _{dc}^b\Omega _{a0}^d  \label{ec.96}
\end{equation}
and

\begin{equation}
{\cal R}_{abc}^d=\partial _b\Omega _{ac}^d-\partial _c\Omega _{ab}^d+\Omega
_{0b}^d\Omega _{ac}^0+\Omega _{eb}^d\Omega _{ac}^e-\Omega _{0c}^d\Omega
_{ab}^0-\Omega _{ec}^d\Omega _{ab}^e.  \label{ec.97}
\end{equation}
Considering (90)-(93) it is straightforward to see that these formulae are
reduced to

\begin{equation}
{\cal R}_{abc}^0=-\partial _bg_{ac}+\partial _cg_{ab}-g_{db}\Gamma
_{ac}^d-g_{db}T_{ac}^d+g_{dc}\Gamma _{ab}^d+g_{dc}T_{ab}^d,  \label{ec.98}
\end{equation}

\begin{equation}
{\cal R}_{a0c}^0=-\partial _0g_{ac},  \label{ec.99}
\end{equation}

\begin{equation}
{\cal R}_{a0c}^b=\partial _0\Omega _{ac}^b  \label{ec.100}
\end{equation}
and

\begin{equation}
{\cal R}_{abc}^d=R_{abc}^d-\delta _b^dg_{ac}+\delta
_c^dg_{ab}+D_bT_{ac}^d-D_cT_{ab}^d+T_{eb}^dT_{ac}^e-T_{ec}^dT_{ab}^e,
\label{ec. 101}
\end{equation}
respectively. Here, we recall that $D_a$ denotes a covariant derivative in
terms of $\Gamma _{ac}^d.$ The Equation (86) implies $\partial _0g_{ac}=0$
and $\partial _0\Omega _{ac}^b=0,$ that is , $g_{ac}$ and $\Omega _{ac}^b$
are independents of $x^0.$ This result remind us the dimensional reduction
procedure in Kaluza-Klein theory$.$

Let us now focus in (98). Since $T_{ac}^d$ is completely antisymmetric,
using (86) the equation (98) leads to

\begin{equation}
\partial _bg_{ac}-\partial _cg_{ab}+g_{db}\Gamma _{ac}^d-g_{dc}\Gamma
_{ab}^d=2T_{cab},  \label{ec.102}
\end{equation}
Combining the indices in (102) we also get

\begin{equation}
\partial _ag_{bc}-\partial _cg_{ba}+g_{da}\Gamma _{bc}^d-g_{dc}\Gamma
_{ba}^d=2T_{cba},  \label{ec.103}
\end{equation}
Thus, adding these two expressions we obtain the equation

\begin{equation}
\partial _bg_{ac}+\partial _ag_{bc}-2\partial _cg_{ab}+g_{db}\Gamma
_{ac}^d+g_{da}\Gamma _{bc}^d-2g_{dc}\Gamma _{ab}^d=0,  \label{ec.104}
\end{equation}
whose solution is

\begin{equation}
\Gamma _{cab}=\frac 12(\partial _ag_{bc}+\partial _bg_{ac}-\partial
_cg_{ab}).  \label{ec.105}
\end{equation}
We recognize in this expression the traditional definition of Christoffel
symbols. Moreover, it is well known that this expression is equivalent to
the metricity condition

\begin{equation}
D_ag_{bc}=0,  \label{ec.106}
\end{equation}
Therefore, we have shown that the metricity condition follows from the
equation (98).

Consider now the expression (101). Using (86) we get

\begin{equation}
R_{abc}^d-\delta _b^dg_{ac}+\delta
_c^dg_{ab}+D_bT_{ac}^d-D_cT_{ab}^d+T_{eb}^dT_{ac}^e-T_{ec}^dT_{ab}^e=0.
\label{ec.107}
\end{equation}
For a homogenous space we have

\begin{equation}
R_{\quad abc}^d=\gamma (\delta _b^dg_{ac}-\delta _c^dg_{ab}),  \label{ec.108}
\end{equation}
where $\gamma $ is a constant. Thus, introducing a new constant $\gamma
^{\prime }=\gamma -1$ the equation (107) becomes

\begin{equation}
\gamma ^{\prime }(\delta _b^dg_{ac}-\delta
_c^dg_{ab})+D_bT_{ac}^d-D_cT_{ab}^d+T_{eb}^dT_{ac}^e-T_{ec}^dT_{ab}^e=0.
\label{ec.109}
\end{equation}
We recognize this expression as the equation (57). Hence, it is
straightforward to prove that expression (109) implies the Cartan-Shouten
equations. Therefore, we have shown that the metricity condition (106) and
the Cartan-Shouten equations follow from (86)-(88).

\smallskip\ 

\noindent {\bf VI. COMMENTS}

\smallskip\ 

It is known that Hurwitz theorem is closely related to the generalized
Frobenius theorem (see ref. 12 and references there in): {\it Every
alternative division algebra is isomorphic to one of the following : the
algebra of real numbers, the algebra of complex numbers, the quaternions,
and the Cayley numbers. }In fact, using Hurwitz theorem the generalized
Frobenius theorem can be proved . Therefore, our procedure also gives an
alternative proof of such a generalized theorem. Let just show how our
procedure can be used to clarify such a relation.

Alternative algebras can be defined by means of the associator

\begin{equation}
(e_i,e_j,e_k)=(e_ie_j)e_k-e_i(e_je_k)\equiv {\cal F}_{ijk}^le_l.
\label{ec. 110}
\end{equation}
In fact, if ${\cal F}_{ijkl}=$ $\delta _{lm}{\cal F}_{ijk}^m$ is completely
antisymmetric for exchanges of any two indices then the algebra is called
alternative. Using (4) and (6) one can show that (86) is equivalent to

\begin{equation}
{\cal F}_{ijkl}=C_{ij}^mC_{mkl}-C_{jk}^mC_{iml}.  \label{ec. 111}
\end{equation}
Now, in section II we showed that normed algebra with an identity implies
that $C_{0j}^m=\delta _j^m,C_{j0}^m=\delta _j^m$ and $C_{ab}^0=-\delta _{ab}$
and that $C_{ab}^c$ is a completely antisymmetric quantity satisfying (32).
From these conditions it follows that

\begin{equation}
{\cal F}_{abcd}=C_{ab}^mC_{mcd}-C_{bc}^mC_{amd}.  \label{ec. 112}
\end{equation}
are the only non-vanishing components of ${\cal F}_{ijkl}.$ Using (32), it
is not difficult to show that this expression leads to

\begin{equation}
{\cal F}_{abcd}=2\{C_{ab}^eC_{cd}^f\delta _{ef}-(\delta _{ac}\delta
_{bd}-\delta _{ad}\delta _{cb})\}=2F_{abcd},  \label{ec. 113}
\end{equation}
where $F_{abcd}$ has been defined in (68). In section IV, we proved that
(113) can be obtained from any normed algebra. Therefore since (111) is
equivalent to (113) we have shown that a normed algebra with an identity is
alternative algebra. The fact that a normed algebra is a division algebra
can be proved directly from the composition law $<AB\mid AB>=<A\mid A><B\mid
B>.$ Indeed, if $AB=0$ the composition law implies that $<A\mid A>=0$ or $%
<B\mid B>=0,$ which means that $A=0$ or $B=0.$ Thus, our procedure based in
tensor analysis gives a straightforward proof of the fact that a normed
algebra with an identity is an alternative division algebra.

It may be interesting to apply the procedure presented in this paper in
different contexts. For instance, it may be helpful to through some light on
the Blencowe-Duff conjecture$^4$: Do the four forces in Nature correspond to
the four division algebras? In fact, part of the motivation of this work
arose as an effort for answering this question. It is known$^{20}$ that
using an algebraic topology called K-theory$^{21}$ we find that the only
dimensions for division algebras structures on Euclidean spaces are again 1,
2, 4, and 8. Therefore, it may be also interesting to relate the present
work to K-theory. Moreover, it is known that Englert's solution of eleven
dimensional supergravity achieves the riemannian curvature-less but
torsion-full Cartan geometries of absolute parallelism on S$^7$. Therefore,
it may be interesting to see if the present work may shed some light to
clarify some aspects of eleven dimensional supergravity which, as it is
known, is the low energy limit theory of M-theory$^{22-27}$. It also seems
interesting to see if tensor analysis may be useful to study the zero
divisors of Cayley-Dickson algebras$^{28}$and Hopf maps. Let briefly outline
this last possibility.

The Cayley-Dickson algebras are defined by the product

\begin{equation}
AB=(A_1B_1-\bar{A}_2B_2,B_2A_1+A_2\bar{B}_1),  \label{ec. 114}
\end{equation}
where $A=(A_1,A_2)$ and $B=(B_1,B_2)$ are in R$^{2^n}=$R$^{2^{n-1}}\times $ R%
$^{2^{n-1}}$ and $\bar{A}=(\bar{A}_1,-A_2).$ Let us denote an algebra with
this structure by ${\it A}_n$. It is found that ${\it A}_0=$real numbers $R$%
, ${\it A}_1=$ complex numbers, ${\it A}_2$ =quaternions and ${\it A}_3=$
octonions. A Hopf map is defined as

\begin{equation}
F_n:{\it A}_n\times {\it A}_n\rightarrow {\it A}_n\times {\it A}_o
\label{ec. 115}
\end{equation}

\[
F_n=(2AB,<B\mid B>-<A\mid A>). 
\]
Consider the multiplication table

\begin{equation}
e_ie_j=D_{ij}^\alpha e_\alpha ,  \label{ec. 116}
\end{equation}
where $D_{ij}^\alpha $ are the structure constants, with $%
i,j,k=0,1,...,2^n-1 $ and $\alpha ,\beta =0,1,...,2^n$. Suppose $%
D_{ij}^\alpha $ satisfies the conditions

\begin{equation}
D_{ij}^{2^n}=\delta _{ij}  \label{ec . 117}
\end{equation}
and

\begin{equation}
D_{ij}^k=-D_{ji}^k,  \label{ec. 118}
\end{equation}
where in (117) we set $\alpha =2^n$. Now, take $H^i=B^i+A^i$ and $%
G^i=B^i-A^i $ and consider the product

\begin{equation}
F^\alpha =H^iG^jD_{ij}^\alpha .  \label{ec. 119}
\end{equation}
Using (117) and (118) we find

\begin{equation}
F^{2^n}=<B\mid B>-<A\mid A>  \label{ec. 120}
\end{equation}
and

\begin{equation}
F^k=2A^iB^jD_{ij}^k.  \label{ec. 121}
\end{equation}
Therefore, $F^\alpha $ defined in (119) reproduces the Hopf map. It remains
to find the relation between $D_{ij}^k$ and the Cayley-Dickson product. At
this respect, our final goal is to see if our procedure may shed some light
on the Hopf maps

\begin{equation}
\begin{array}{c}
F_0:S^1\rightarrow S^1, \\ 
\\ 
F_1:S^3\rightarrow S^2, \\ 
\\ 
F_2:S^7\rightarrow S^4, \\ 
\\ 
F_3:S^{15}\rightarrow S^8.
\end{array}
\label{122}
\end{equation}
which have a certain topological invariant , the Hopf invariant, equal to
one.

Finally, it may be interesting to find the connection between the present
paper and the Wolf`s works of references 29 and 30, in which the
Cartan-Shouten formalism is generalized to the case of non-Euclidean spaces.
Moreover, a possible connection between our procedure and flexible
Malcev-admissible algebras (see references 31, 32 and 33 and references
there in) may deserve further research.

\smallskip\


\begin{references}
\bibitem{1}  T.$\frac{{}}{{}}$ Kugo and P. Townsend, Nucl. Phys. {\bf B221,}
357(1983).

\bibitem{2}  J. M. Evans, Nucl. Phys. {\bf B298,} 92 (1988).

\bibitem{3}  G. Sierra, Class. Quantum Grav.{\bf \ 4, }227 (1987).

\bibitem{4}  M. P. Blencowe and M. J. Duff,{\it \ }Nucl. Phys. {\bf B310,}
387 (1988).

\bibitem{5}  J. M. Figueroa-O'Farril, J. Geom. Phys. {\bf 32}, 227 (1999).

\bibitem{6}  T. Kimura, I. Oda, Prog. Theor. Phys. {\bf 80}, 1 (1988).

\bibitem{7}  I. Oda, T. Kimura and A. Nakamura, Prog. Theor. Phys. {\bf 80},
367 (1988).

\bibitem{8}  I. Bengtsson, Nucl. Phys. {\bf B302}, 81 (1988).

\bibitem{9}  R. Foot and G. C. Joshi, Lett. Math. Phys. {\bf 16}, 77 (1988).

\bibitem{10}  C. A. Monogue and J. Schray, J. Math. Phys. {\bf 34}, 3746
(1993).

\bibitem{11}  C. M. Hull, JHEP, {\bf 9811}, 017 (1998); hep-th/9807127.

\bibitem{12}  I.L. Kantor and A.S. Solodovnikov, {\it Hypercomplex Numbers;
An Elementary Introduction to Algebras (}Springer-Verlag New York, 1989).

\bibitem{13}  S. Okubo, {\it Introduction to Octonion and Other
Non-Associative Algebras} (Cambridge University Press, 1995).

\bibitem{14}  K. Abdel-Khalek, Int. J. Mod. Phys. {\bf A13,} 569 (1998);
hep-th/9704049.

\bibitem{15}  J. Adem, Obra Matem\'{a}tica, El Colegio Nacional, M\'{e}xico,
(1992).

\bibitem{16}  F. R. Cohen, Bol. Soc. Mat. Mex. {\bf 37 }55 (1992).

\bibitem{17}  Y. A. Drozd and V. V. Kirichenko, {\it Finite Dimensional
Algebras} (Springer-Verlag, 1994).

\bibitem{18}  E. Cartan and J. A. Schouten, Proc. K. Akad. Wet. Amsterdam 
{\bf 29,} 803 (1926).

\bibitem{19}  F. Gursey and C. Tze,{\it \ }Phys. Lett. {\bf B127,} 191
(1983).

\bibitem{20}  N. Steenrod, {\it The Topology of Fibre Bundles} (Princeton
University Press, 1970).

\bibitem{21}  M. F. Atiyah. {\it K-theory} (W. A. Benjamin Inc., 1979).

\bibitem{22}  P. K. Townsend, {\it Four Lectures on M Theory}, ICTP Summer
School on High Energy Physics and Cosmology, Trieste, June 1996;
hep-th/9612121.

\bibitem{23}  M. J. Duff, Int. J. Mod. Phys. {\bf A11, }5623{\bf \ }(1996);
hep-th/9608117.

\bibitem{24}  P. Horava and E. Witten, Nucl. Phys. {\bf B460, }506 (1996).

\bibitem{25}  M. J. Duff, R. R. Khuri and J. X. Lu, Phys. Rep. {\bf 259,}
213 (1995).

\bibitem{26}  E. Witten,{\it \ }Nucl. Phys. {\bf B463, }383 (1996).

\bibitem{27}  J. H. Schwarz, Nucl. Phys. Proc. Suppl. {\bf B55, }1 (1997);
hep-th/9607201.

\bibitem{28}  G. Moreno, {\it The zero divisors of Cayley-Dickson algebras,}
Reporte Interno CINVESTAV, No. {\bf 195}, (1995); {\it G}$_{2}$ {\it como
Soluci\'{o}n a un Problema de Teor\'{i}a de Singularidades, }Ap. Mat. Ser.
Com. {\bf 20}, 129 (1997).

\bibitem{29}  J. A: Wolf, J. Diff. Geom. {\bf 6} (1972) 317.

\bibitem{30}  J. A: Wolf, J. Diff. Geom. {\bf 7 }(1972) 19.

\bibitem{31}  H. C. Myung, ${\it Malcev}$ ${\it admissible}$ ${\it algebras}$
(Birkhauser, Boston, 1986).

\bibitem{32}  M. L. El-Mallah, Archiv. der Math. {\bf 49 }16 (1987); and 
{\bf 51} 39 (1988) .

\bibitem{33}  S. Okubo, Hadronic J.{\bf \ 4, }216 (1981).
\end{references}
\end{document}